\documentstyle[12pt]{article}

\newcommand{\be}{\begin{equation}}
\newcommand{\ee}{\end{equation}}

\begin{document}
\sloppy

{\centerline {\bf Comment on }}

{\centerline{\bf " Thermodynamics of a One- Dimensional
Ideal Gas with fractional Exclusion Statistics"}}
\vskip 1cm
{\centerline {\bf \rm Alain DASNI\`ERES de VEIGY and St\'ephane
OUVRY\footnote{\it  and
LPTPE, Tour 12, Universit\'e Paris  6 / electronic e-mail: OUVRY@FRCPN11}}}
{\centerline {Division de Physique Th\'eorique\footnote{\it Unit\a'e de
Recherche des Universit\a'es Paris 11 et Paris 6 associ\a'ee au CNRS},
IPN, Orsay Fr-91406}}

\vskip 1cm


In [1], the conclusion was reached that,
in the one-dimensional
Calogero model,
only the second virial coefficient is affected by the statistical
parameter $\alpha$, where $\alpha$
is related to the coupling constant $\kappa/ x_{ij}^2$ of the Calogero
interaction
by $\kappa=\alpha(\alpha+1)$ ($\alpha\to -g$ in their notation).
We argue
that it is not so, i.e. all virial coefficients are affected,
if the thermodynamic limit is properly
taken. In [1],  the system is regularized at long distance
by harmonically attracting the
particles to the origin, with
 the harmonic pulsation
$\omega\to 0$ in the thermodynamic limit. However, this limit should be taken
with a
particular prescription that will be detailed
below.  In fact, the correct cluster expansion was already conjectured by
Sutherland
in its original paper [2] on the thermodynamic of the Calogero
model, by various means, in particular a harmonic regularization.
On the other hand,
the anyon model in the lowest Landau level (LLL) of an external $B$-field
[3]
 (LLL-anyon model) shares some similarities with the
Calogero model, if the anyonic statistical parameter $\alpha\in[-1,0]$ modulo 2
(by convention, $eB>0$ has been assumed).
The
link with Haldane's statistics Hilbert space counting argument [4] is
immediate, because of the
diminution  of available one-body quantum states in the LLL, due to
the screening of the $B$ field.
 In this regime,
the groundstate is known, excited states have a gap, and the LLL-anyon model
can be viewed as a one-dimensional model.
In the sequel, we will treat simultaneously both models -for a review on
the subject see [5].

In a harmonic well, the $N$-body spectrum of the Calogero or LLL-anyon models
is
$\sum_{\ell}n_\ell\epsilon_{\ell} -{1\over2}N(N-1)\varpi\alpha$,
where, for a given
angular momentum $l$,  the 1-body energy  is
$\epsilon_\ell=\epsilon_0+\ell\varpi$.
The LLL-anyon spectrum corresponds to $\varpi=\omega_{\rm t}
-\omega_{\rm c}$ ($\omega_{\rm t}=\sqrt{\omega_{\rm c}^2+\omega^2}$,
$\omega_{\rm c}=eB/(2m)$) and $\epsilon_0=\omega_{\rm t}$,
whereas the Calogero spectrum corresponds to $\varpi=\omega$
and $\epsilon_0={1\over2}\omega$.
{}From the $N$-body partition function
\be
Z_N^{\omega}=e^{{1\over2}\beta N(N-1)\varpi\alpha}
\prod_{n=1}^N{e^{-\beta\epsilon_0}
\over 1-e^{-n\beta\varpi}},
\ee
one deduces the cluster expansion of the thermodynamical potential
$\Omega^{\omega}\equiv-\ln\sum_NZ_N^{\omega}z^N=-\sum_nb_n^{\omega}z^n$.
In the limit $\omega\to0$
\be  \label{polypoly}
b_n^{\omega}={1\over\beta\varpi} {e^{-n\beta\epsilon_0}\over n^2}
\prod_{k=1}^{n-1}
{k+n\alpha\over k}
\ee

The thermodynamic limit prescription
in (2) should be dictated by the requirement that
when $\omega\to 0$, the thermodynamical potential of a system of
$d$-dimensional harmonic oscillators coincide with those of a system of
$d$-dimensional particles in a box of infinite volume $\Omega^V=-\sum_n
b_n^Vz^n$. Close to $\vec r$,
one can approximate $\sum_i {1\over 2}m\omega^2 {\vec{r}_i}\,^2$
by ${N\over2}m\omega^2{\vec{r}}\,^2$.
It follows that, for a system of particles
confined in an infinitesimal  volume $d^d\vec {r}$ around
$\vec r$,
the local thermodynamical potential
 in the presence of a small harmonic attraction
can be approximated by
replacing
$z\to z e^{-{1\over2}\beta m\omega^2{\vec r}\,^2}$ and $V\to d^d{\vec r}$ in
the
infinite box
thermodynamical potential.
Since the thermodynamical potential is additive in the limit $\omega\to0$,
one has
for the entire system ($\lambda$ is the
thermal wavelength)
\be \Omega^{\omega}=-\int {d^dr\over V} \sum_{n=1}^\infty b_n^V
    \left(z \ e^{-{1\over2}\beta m\omega^2r^2}\right)^n=-\sum_{n=1}^\infty
    {\lambda^d\over n^{d/2}(\beta\omega)^dV}\, b_n^V z^n
\ee
Thus one has $b_n^{\omega}\to b_n^V$, provided that the prescription
\be 1/(n\beta^2\omega^2)^{d/2}\to V/\lambda^d\ee
is enforced at each order $z^n$. One deduces the following results :

{\bf LLL-anyon model : }
The cluster coefficients are
\be b_n^{V}= {V\over \lambda^2}{e^{-n\beta\omega_c}\over n} \prod_{k=1}^{n-1}
{k+n\alpha\over k}\ee
The thermodynamical potential reads
$
\Omega^V=-V\rho_{\rm L}\ln y
$,
where $y$ is solution of
$y-ze^{-\beta \omega_c}y^{1+\alpha}=1$.
This is precisely because of the $d=2$ prescription given in (4)
that the Landau degeneracy $\rho_LV$ has factorized.
The virial coefficients are
$a_n= (-{1\over \rho_{\rm L}})^{n-1} {1\over
n}\{(1+\alpha)^n-\alpha^n\}$.

{\bf Calogero model :}
The cluster coefficients are
\be  \label{100}
 b_n^{V}= {V\over \lambda^2}{1\over n^{3/2}} \prod_{k=1}^{n-1}
{k+n\alpha\over k}
\ee
The thermodynamical potential is
$\Omega^V=-{V\over2\pi}\int dp  \ln y
$,
where $y$ is now solution of
$y-ze^{-\beta p^2/2m}y^{1+\alpha}=1$.
The  continuous one-dimensional
energy spectrum   $p^2/2m$ appears precisely because of the
$d=1$ prescription (4). The virial coefficients are all
$\alpha$ dependent, $ a_2= -\sqrt{2}({1\over 4}+{\alpha\over 2}),
a_3= {1\over 2} +2\alpha+2\alpha^2-\sqrt{3}({2\over
9}+\alpha+\alpha^2)$,
et caetera.
 In [1], the inappropriate thermodynamic limit prescription
$1/(\beta^2\omega^2)^{d/2}\to V/\lambda^d$ was
taken, and a virial expansion interpolating
between the $d=2$ Bose and Fermi gases was obtained, where only
the second virial coefficient is affected by $\alpha$. It would be
certainly satisfactory to find a $d=2$ $N$-body Hamiltonian which would lead to
such a simple virial expansion.

\end{document}